\documentclass[fleqn,usenatbib]{mnras}

\usepackage{newtxtext,newtxmath}
\usepackage{xcolor}
\usepackage{ulem}
\usepackage{color}

\newcommand{\trotta}[1]{{#1}}

\usepackage{lineno}
\usepackage[T1]{fontenc}

\DeclareRobustCommand{\VAN}[3]{#2}
\let\VANthebibliography\thebibliography
\def\thebibliography{\DeclareRobustCommand{\VAN}[3]{##3}\VANthebibliography}

\usepackage{graphicx}	
\usepackage{amsmath}	

\newcommand{\tbn}{$\theta_{Bn}$}	

\title[Ion diffusion in turbulent shocks]{A study of the transition to a turbulent shock using a coarse-graining approach to ion phase space transport.}

\author[D. Trotta et al.]{D. Trotta,$^{1,2}$\thanks{E-mail: d.trotta@imperial.ac.uk }
F. Valentini$^{3}$,
D. Burgess$^{4}$,
S. Servidio$^{3}$
\\
$^{1}$ European Space Agency (ESA), European Space Astronomy Centre (ESAC), Camino Bajo del Castillo s/n, 28692 Villanueva de la Cañada, Madrid, Spain \\
$^{2}$ The Blackett Laboratory, Department of Physics, Imperial College London, London SW7 2AZ, UK \\
$^{3}$Dipartimento di Fisica, Universit\`a della Calabria, Rende 87036, Italy \\
$^{4}$ Department of Physics and Astronomy, Queen Mary University of London, E1 4NS, UK \\
}

\date{Accepted XXX. Received YYY; in original form ZZZ}

\pubyear{2023}

\begin{document}
\label{firstpage}
\pagerange{\pageref{firstpage}--\pageref{lastpage}}
\maketitle

\begin{abstract}
	
Shocks and turbulence are ubiquitous phenomena, responsible for particle acceleration to very high energies in a large collection of astrophysical systems. Using self-consistent, hybrid-kinetic simulations with and without pre-existing turbulence, we study the transition of a shock from ``laminar'' to turbulent. We show that the changes in upstream proton transport behaviour are crucial to understand this transition, which we address quantitatively with a novel Eulerian approach. This method, based on the coarse-graining of the Vlasov equation originally introduced in one of our previous studies, gives consistent results for inertial range scales. The potential applications of the coarse-graining approach beyond the shock-turbulence system are outlined. 

\end{abstract}




\section{Introduction}
\label{sec:introduction}

Energetic particles are omnipresent in our universe, and the mechanisms for their acceleration are not completely 
clear \citep[e.g.,][]{Bykov2019,Amato2017}. Shocks, which are commonly present in our universe, are well-known efficient 
particle accelerators, as they convert directed flow energy (upstream) to thermal and magnetic energy (downstream) 
and, in the collisionless case, a fraction of the available energy can be channeled in the production of energetic particles 
\citep[e.g.,][]{Blandford1987,Fisk1974}. Another important source of accelerated particles is plasma 
turbulence, a pervasive state of astrophysical systems \citep[e.g.,][]{Lazarian2012,Petrosian2012}. Crucially, shock waves 
and plasma turbulence can be directly observed in the heliosphere by means of spacecraft measurements, thus providing the missing link to remote astrophysical systems~\citep{Burgess2015}.

An important parameter that strongly influences the behaviour of collisionless shock transitions is the angle between the
normal to the shock surface and the upstream magnetic field, \tbn{}. When \tbn{} is close to zero (and hence the upstream 
magnetic field is almost normal to the shock surface), the shocks are classified as quasi-parallel, and they are believed to be 
efficient proton accelerators \citep[e.g.,][]{Giacalone1997,Reville2013,Caprioli2014a}. When \tbn{} is around 90$^\circ$, 
shocks are dubbed quasi-perpendicular and efficiently accelerate electrons
\citep[e.g.,][]{Leroy1984,Wu1984,Kraussvarban1989a,Trotta2019, Matsumoto2017}. Thus, the local shock geometry and its fluctuations are fundamental for shaping how particles are
accelerated to high energies, as also shown in recent observations of Earth's bow shock resulting from the interaction between the supersonic solar wind and Earth's magnetosphere~\citep[][]{Lindberg2023} and interplanetary shocks due to solar activity phenomena~\citep[][]{Trotta2023b}.

Oblique shocks ($40^\circ \lesssim \theta_{Bn} \lesssim 75^\circ$) have an intermediate behavior between the two discussed 
above. A prominent feature of oblique shocks is the presence of upstream, counter-propagating ion Field Aligned Beams (FABs) 
\citep[][]{Gosling1978,Paschmann1980,Schwartz1983}. Observations of FABs at Earth's bow shock revealed a collimated 
backstreaming ion population with energies of a few keV, in contrast with the diffuse, energetic ion population originating at 
the quasi-parallel portion of the bow shock \citep[e.g.,][]{Bonifazi1980}. 

The mechanisms leading to the production and propagation of FABs are still subject to many open questions. 
For example, the key idea for production of FABs is that they originate due to pitch angle scattering in the reflected ion 
population in the oblique shock ramp \citep[][]{Mobius2001,Kucharek2004}, but the source of such scattering still 
needs to be identified. From this point of view, the fact that shocks are often propagating in turbulent media 
is of key importance to understand the production and transport properties of FABs and to shape their morphology in velocity space, from crescent-like distributions to more irregular ones~\citep{Kis2007}, as addressed in this work. Recently, \citet{Lario2022} reported on the presence of very long-lasting FABs upstream of interplanetary shocks, which were put in the context of a low level of upstream fluctuations, thus indicating that the environment where shocks propagate plays a fundamental role in the injection mechanism of energetic particles.

Turbulence is ubiquitous in astrophysical systems, 
and widely accepted as another key mechanism leading to energetic particles
\citep[e.g.,][]{Jokipii1966, Matthaeus2003,Ambrosiano1988}. 
In the past decade, the role of coherent structures and 
current sheets, naturally arising from turbulent systems,  has been proven to be of key importance on the particle acceleration and transport mechanisms~\citep[e.g.,][]{Arzner2004,Dmitruk2004,Servidio2016,Trotta2020b,Comisso2018}. 

The interaction and reciprocal influence of shock and turbulence is of pivotal importance to understand the mechanisms of 
particle energisation and transport, as discussed in early theoretical and numerical 
works and as observed in heliospheric environments~\citep[see][for a review]{Guo2021}. From the numerical perspective, \citet{Giacalone2005b} considered the problem of shock accelerated particles
including pre-existing magnetic field irregularities using a MagnetoHydroDynamic (MHD) approach combined with test-particles. 
The acceleration of electrons at shocks moving through turbulent fields was also investigated, and efficient electron 
acceleration was found for a broad range of shock geometries \citep[][]{Guo2012,Guo2015}. 
Shock dynamics in presence of current sheet and turbulence has been rcently considered in 
\citet{Nakanotani2021,Nakanotani2022}, elucidating how downstream particle energy spectra are influenced by turbulence.

Analogously, \citet{Zank2002} formalised self-consistently the interaction of turbulence and shocks with a fluid treatment, 
and found that the mean shock speed increases with increasing levels of upstream turbulence, implying that turbulent 
fluctuations are not simply transmitted and amplified across the shock, but are also converted in mean (downstream) flow 
energy. An improvement to such effort including the magnetic field was recently carried out in~\citet{Gedalin2023}.
The interplay between acceleration at coherent structures and diffusive shock acceleration (DSA) has been considered by \citet{Zank2015} using a transport 
formalism developed in \citet{Zank2014}, concluding that both the shock transition and the coherent structures interacting 
with it are important ingredients for particle acceleration.

More recently, using 
a combination of MHD and hybrid kinetic simulations, \citet{Trotta2021} modelled the interaction between oblique shocks and
fully-developed turbulence elucidating that particle diffusion in phase space is enhanced. The same approach was used to model 
the interaction between turbulent structures and perpendicular shocks in both reduced two-dimensional and full three-
dimensional geometries~\citep{Trotta2022a,Trotta2023c}.

\trotta{Coarse-graining techniques are widely utilized in the investigation of complex systems, particularly in the context of turbulence, where nonlinearities generate finer scales, encompassing phenomena from Navier-Stokes~\citep{Germano1992,Eyink2009} to Magnetohydrodynamics (MHD)~\citep{Aluie2017,Yang2016}. The approach of filtering through smooth coarse graining of the fields is a standard practice in both large-scale simulations and analyses of turbulent fields. Germano-like techniques are consistently employed to model unresolved small-scale physics, applicable in principle to any complex system. The Vlasov equations certainly fall within this category of nonlinear systems, as they exhibit the development of small-scale features, as documented in our research and all prior kinetic simulations of turbulence\citep{Eyink2018}. However, the coarse graining method for plasmas remains at a nascent stage when compared with fluid models.}
Motivated by the above studies, in this paper, we 
address particle transport in the transition from laminar to turbulent shock, performing simulations with increasingly strong pre-existing turbulence presented originally in \citet{Trotta2021}. We report on how shock transitions, downstream energy spectra and suprathermal particles change with different upstream turbulence conditions. To understand the transition to turbulent shock, we lay the theoretical and numerical framework for the novel, Eulerian approach relying on the
coarse-graining of the Vlasov equation. We then show how the particles' transport properties change at different spatial/velocity scales and demonstrate that our new model gives consistent results if the range of the filtering scales belongs to the inertial range of turbulence.

The paper is organised as follows. In Section \ref{sec:simulations}, we give the details about the simulations method. In 
Section \ref{sec:shocks_particles}, we give an overview of the perturbed shocks and energetic particle behaviour. In 
Section~\ref{sec:particles_transport}, the coarse-grained approach is introduced from a theoretical perspective and then applied to the shock-turbulence simulations in Section \ref{subsec:CG_applied}. The conclusions are given in 
Section \ref{sec:conclusions}.

\section{Numerical Simulations}
\label{sec:simulations}
Our simulations method consists of two stages. First, plasma turbulence is generated by means of compressible MHD simulations. \trotta{The resulting turbulent fields are then injected in a hybrid Particle-In-Cell (PIC) shock simulation, obtaining a shock that is generated and propagates in a turbulent upstream.}

Fully developed, decaying, homogeneous plasma turbulence is first produced via 2.5D compressible MHD simulations \citep[see][ for further details]{Perri2017}. \trotta{In such simulations, initial fluctuations at large scales, initialised with random phases, trigger a turbulent cascade.}
Three simulations have been performed, with different levels of turbulence fluctuations, namely $\delta B/B_0 = 0.0, 0.4, 0.8, 2.1$, where $B_0$ is the mean field, at $\theta_{Bn} = 45^\circ$ in the $x$-$y$ plane, and $\delta B$ is the level of fluctuations, measured when the MHD simulation reaches a steady state. \trotta{Note that the dB/B = 2.1 case represents an extreme choice for heliospheric shocks, and it is here discussed to show a very strongly perturbed shock. However, such conditions may be observed in some CME events~\citep{Good2020}. Further, such strongly perturbed cases may be extremely relevant for astrophysical systems where the mean magnetic field is very low, such as intracluster medium shocks~\citep[e.g.,][]{Kang2019}.} When the turbulence is fully developed and coherent structures form, the output is stored to be used as an initial condition for the shock simulation. In the MHD simulations, typical Alfv\`en units have been used, on a periodic box of generic side $L_0$.

To use the output from the MHD simulations as the initial condition for the shock simulation, an additional step is necessary. This is because the windowed MHD simulations are performed on a double-periodic domain, whereas the shock simulation features open boundary-reflecting walls along the $x$-axis. Therefore, the vector potential obtained from the MHD simulation is utilized.

The vector potential is set to zero in real space at the boundaries of the simulation domain using the following ``mask'' function:
\begin{align}
	f(x) =  \frac{1}{2}\left[1+\tanh\left(\frac{x-\sigma_1}{\delta}\right)\right]  \times \nonumber \\ 
	\times \frac{1}{2}\left[1-\tanh\left(\frac{x -\sigma_2}{\delta}\right)\right],
\end{align}
where $\sigma_1$ and $\sigma_2$ control the points in space at which the filter is activated and the $\delta$ parameter controls the range over which the function goes to zero.

\begin{figure}
\includegraphics[width=0.48\textwidth]{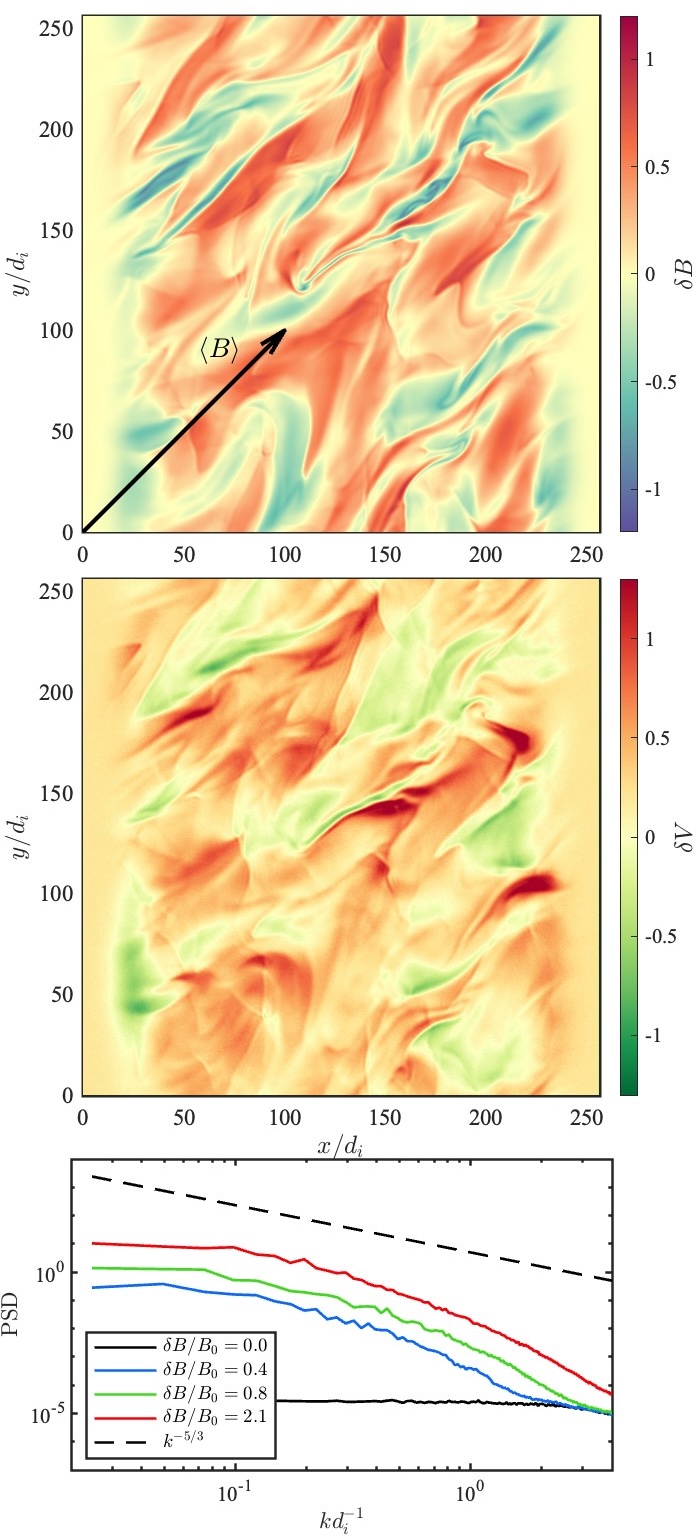}
\caption{\trotta{Top two panels: Two-dimensional panels showing the initial conditions for the moderately perturbed case ($\delta B/B_0 = 0.8$). In the top and middle panels are shown, respectively, the magnitude of magnetic field and ion bulk flow speed perturbations $\delta B$ and $\delta V$. Bottom panel: one-dimensional magnetic field power spectral density (PSD) for all the simulation cases presented. }
\label{fig:inicond}}
\end{figure}

 \trotta{Initial particle speeds are also perturbed for the kinetic simulations, by using the turbulent MHD fields. In particular, in each cell of the simulation domain, the proton velocity distribution function (VDF) is shifted in velocity space by an extent $\delta V$, corresponding to the value obtained by the MHD simulations. In practice, this corresponds to modifying, cell by cell,  the initial speed of each (macro) particle by $\delta V$.}
To be consistent with the filtering method used for the magnetic vector potential, a Helmholtz decomposition has been applied to the MHD velocity field. The decomposition is performed such that $\mathbf{U} = \mathbf{\nabla}\psi \times \mathbf{\hat{z}} - \mathbf{\nabla} \phi$, where $\psi$ and $\phi$ are the velocity potentials. By inverting the above equation with spectral techniques, we extract the potentials, that we filter similarly to the magnetic potential. With this procedure, we finally obtain the windowed velocity field. \trotta{Figure~\ref{fig:inicond} shows the initial condition for the perturbed shock simulations. The top two panels show the spatial pattern of the magnitude of magnetic field and ion bulk velocity perturbations $\delta B$ and $\delta V$. It can be seen that the typical turbulent spatial pattern observed in simulations with an in-plane mean magnetic field  is imposed~\citep{Camporeale2011}. The action of the windowing is also evident, with the perturbation amplitudes going to zero at the left- and right-hand side boundaries of the simulation in the $x$-direction. Further, the bottom panel of Figure~\ref{fig:inicond} shows the magnetic field  power spectral density (PSD) for the unperturbed (black) and perturbed cases, highlighting how stronger turbulence levels are chosen throughout the simulation campaign. The PSDs are computed  in the $y$- (periodic) direction, then  integrated over the nominal shock normal direction. Spectral laws typical of plasma turbulence are recovered, and form the initial condition for the simulation setup.}

The second (main) stage of the simulation consists in running the (perturbed) shock simulations. The HYPSI code has been used for this purpose \citep[e.g.,][]{Trotta2019}. Here, protons are modelled as macroparticles and advanced using the standard PIC method. The electrons, on the other hand, are modelled as a massless, charge-neutralizing fluid with an adiabatic equation of state with an adiabatic index of 5/3. The HYPSI code is based on the CAM-CL algorithm \citep[][]{Matthews1994},used also in other PIC codes~\citep{Franci2015}.

\begin{figure*}
\includegraphics[width=.99\textwidth]{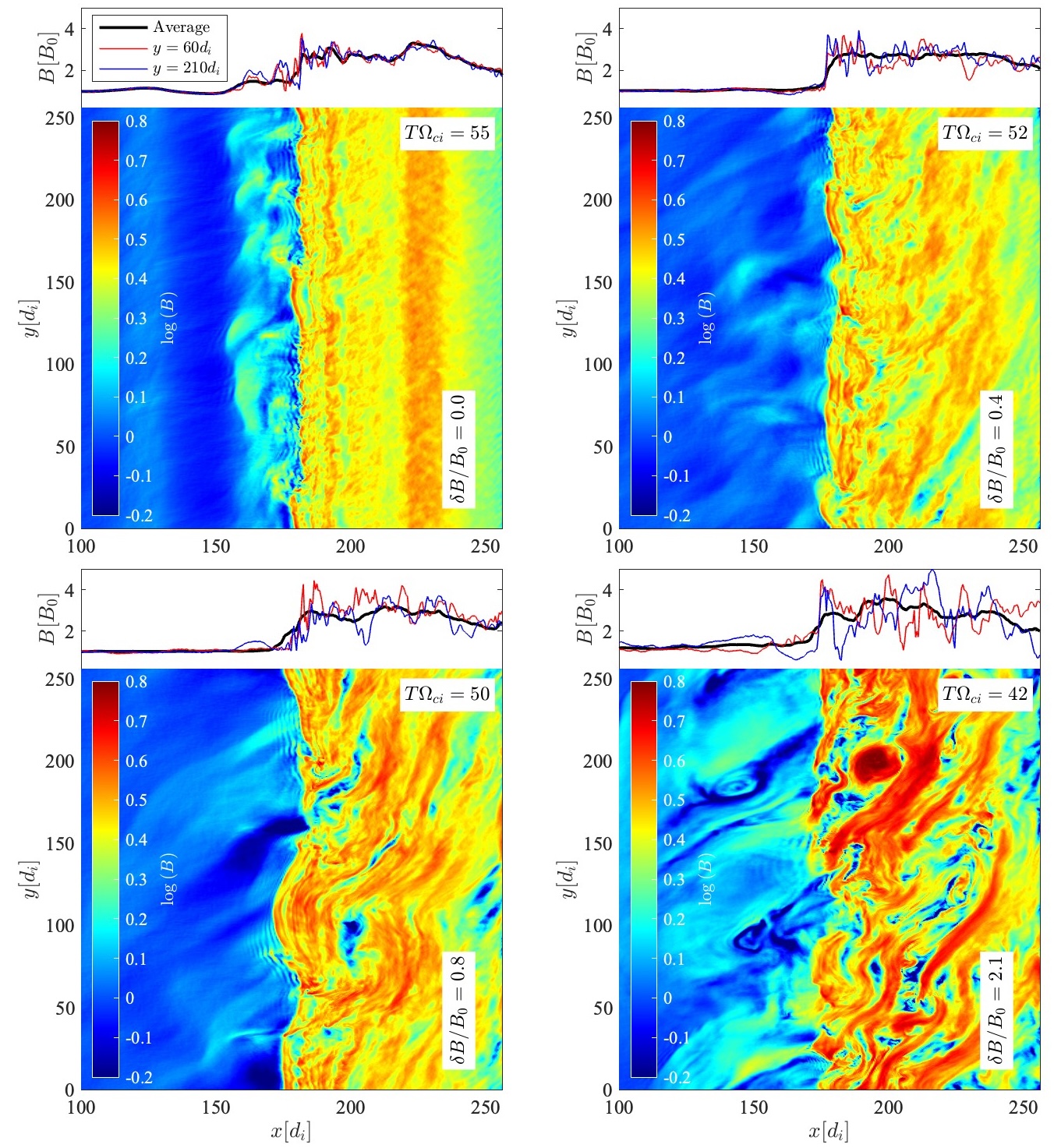}
\caption{\trotta{Magnetic field magnitude snapshots for all simulation cases, where the nominal shock position is around 170 $d_i$. It may be noted that the snapshots are taken at different times due to the slightly different shock speeds. The colormaps show the magnetic field intensity in the two-dimensional simulation domain. In the upper panels, we show the $y$-averaged magnetic field profile (black line) as well as along two different horizontal cuts at $y=60 d_i$ and $y=210 d_i$ (red and blue lines).} \label{fig:overview}}
\end{figure*}

The shock is initiated by the injection method \citep[][]{Quest1985}, in which the plasma flows in the $x$-direction with a defined (super-Alfv\'enic) velocity $V_\mathrm{in}$. The right-hand boundary of the simulation domain acts as a reflecting wall, and at the left-hand boundary plasma is continuously injected. The simulation is periodic in the $y$ direction. A shock is created as a consequence of reflection at the wall, and it propagates in the negative $x$-direction. In the simulation frame, the (mean) upstream flow is along the shock normal.

In the hybrid simulations, distances are normalised to the ion (proton) inertial length $d_i \equiv c/\omega_{pi}$, times to the inverse cyclotron frequency ${\Omega_{ci}}^{-1}$, velocity to the Alfv\'en speed $v_A$ (all referred to the unperturbed upstream state), and the magnetic field and density to their unperturbed upstream values, $B_0$ and $n_0$, respectively. For the upstream flow velocity, the value $V_\mathrm{in} = 3.5 v_A$  has been chosen, and the resulting Alfv\'enic Mach number of the shock is approximately $M_A = 6$. The upstream ion distribution function is an isotropic Maxwellian and the ion $\beta_i$ is 1. \trotta{The electron $\beta_e$ is also set to 1}.  The simulation $x-y$ domain  is 256 $\times$ 256 $d_i$. The spatial resolution used is $\Delta x$ = $\Delta y$ = 0.5 $d_i$. \trotta{Such choice of resolution enables us to study particle dynamics in the shock system 
and its surroundings, compatible with previous studies with both heliospheric and 
astrophysical parameters~\citep[e.g.,][]{Burgess20163d,Caprioli2014, Boula2024}, while 
to retain the smallest scales involved in the shock transition, a fully kinetic approach may
be necessary, though suffering from other computational 
limitations~\citep{Riquelme2011}.}The final time for the simulation is 90 $\Omega_{ci}^{-1}$, the time step for particle {(ion)} advance is $\Delta t_{}$  = 0.01 $\Omega_{ci}^{-1}$. Substepping is used for the magnetic field advance, with an effective time step of $\Delta t_{B} = \Delta t_{}/10$. A small, nonzero  resistivity is introduced in the magnetic induction equation. The value of the resistivity is set to \trotta{0.01 $\omega_p^{-1}$}, chosen so that is small but there are not excessive fluctuations at the grid scale. \trotta{We initialise the simulations with 500 particles per cell (upstream), in order to keep the statistical noise characteristic of PIC simulations to a reasonable level. Note that new particles are injected at each time at the open boundary at the $x$=0 open boundary}.

\begin{figure*}
\includegraphics[width=.99\textwidth]{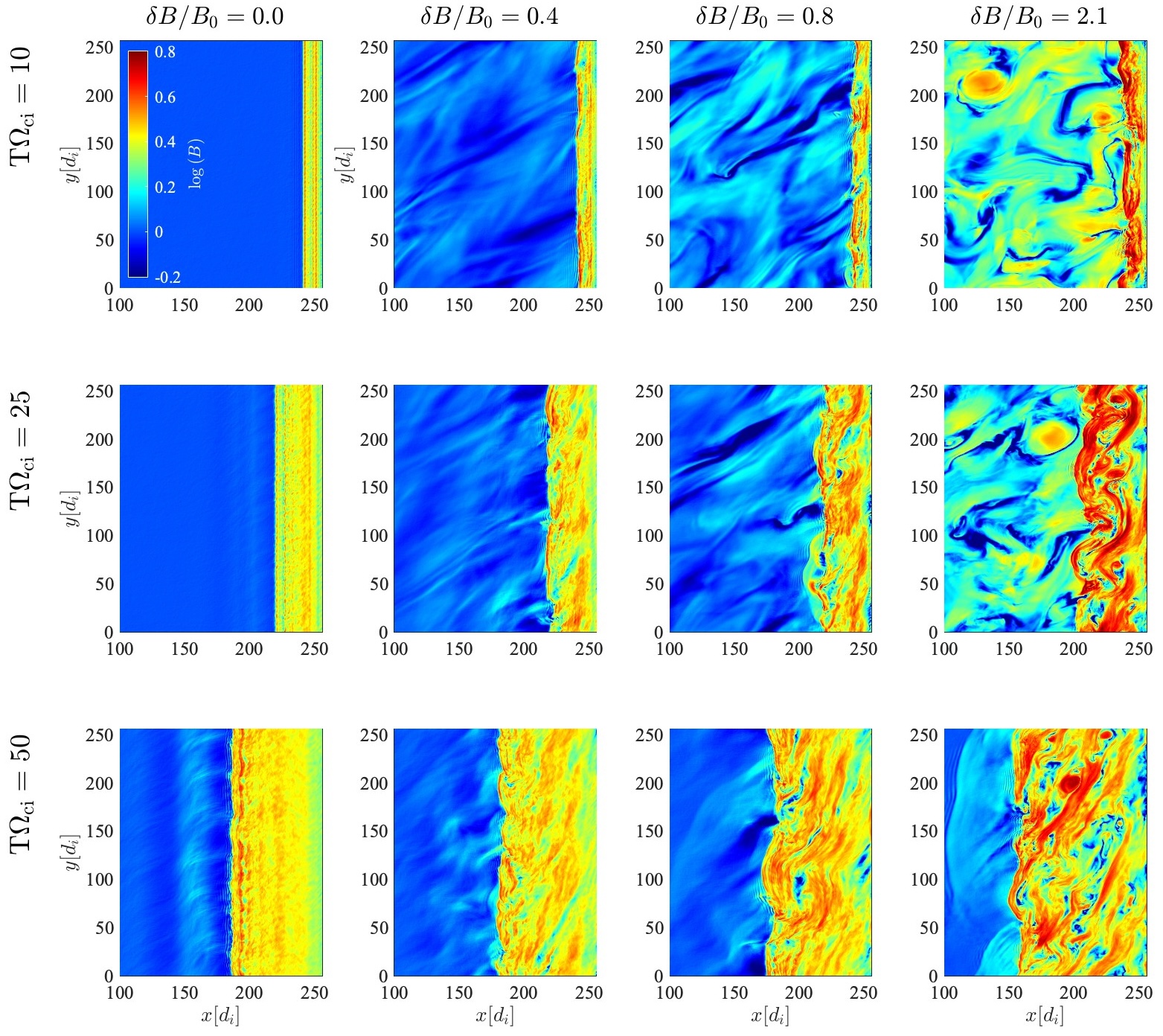}
\caption{\trotta{Magnetic field magnitude snapshots for all simulation cases (coloumns showing the $dB/B_0$ = 0, 0.4, 0.8 and 2.1 cases, respectively)) and for three different simulation cases (rows, showing simulation times $\mathrm T \Omega_{ci} = 10,25,50$, from top to bottom), showing the different evolution stages of the shock front.} \label{fig:evolution}}
\end{figure*}

An important aspect of this initialisation is that the perturbation is finite in both space and time extent. In fact, once the ``turbulent patch'' starts to be convected with the upstream flow, at the injection boundary, new, unperturbed plasma is introduced in the simulation. On one hand, this represent a limitation, as the time of interaction between the shock front and the turbulent perturbation is only a fraction of the total simulation time. On the other hand, this strategy may be relevant for modelling shocks interacting with strongly inhomogneous environments. Note that our procedure is very advantageous from the numerical point of view, since it provides, at the upstream region, genuinely generated MHD turbulence, with coherent structures such as vortices, islands and MHD waves. This field is more realistic than arbitrary random phases perturbations.

\section{Perturbed shocks and energetic particles}
\label{sec:shocks_particles}

We start by describing the overall features of the shock in its transition from laminar to turbulent. An overview of the shock simulations performed, from the unperturbed ($\delta B/B_0=0$) to the highly turbulent case ($\delta B/B_0=2.1$) is shown in Figure \ref{fig:overview}. \trotta{In this analysis, like in the rest of the manuscript, the unperturbed case is compared with the perturbed cases with increasing levels of turbulence.} \trotta{We plot four simulation snapshots when the shock front is well-developed and not influenced anymore by the reflecting wall at the right-hand side of the simulation box. For all cases, the nominal shock position is at about 170 $d_i$.} \trotta{Note that each of these snapshots corresponds to different simulation times due to the fact that the shock has a slightly different speed in the four cases, as discussed below}. The shocks have the same nominal upstream parameters, and the turbulence level is increased from case to case. 

The magnetic field intensity maps in Figure \ref{fig:overview} give an 
overview of the spatial behaviour of the shocks. When no upstream perturbations are present ($\delta B/B_0 = 0.0$), 
the shock has the typical structure of oblique, supercritical shocks. Magnetic fluctuations can be seen upstream, due to the
presence of a FAB of counterstreaming protons, as extensively investigated in previous literature 
\citep[e.g.][]{Preisser2020,Young2020}. Finally, the shock downstream has moderate to strong fluctuations in the vicinity of 
the shock ramp, with a relaxation further away from it.

As it can be seen from Figure \ref{fig:overview} that, even with a low level o pre-existing turbulence, the situation changes dramatically. The first important observation has to do with the shock transition itself, that is shown to be much more irregular, 
unstable and undulated, due to the disturbance generated by the upstream turbulence. The observation of this feature being already present in 
the ``weak'' upstream turbulence case ($\delta B/B_0 = 0.4$) are compatible with the predictions of 
\citet{Zank2002}, where possible destabilisations of the shock front even with low amplitude turbulence are discussed. The shock front distorsions observed here are compatible with the
ones reported for turbulence--mediated perpendicular shocks by \citet{Nakanotani2022}.
Furthermore, a complete characterisation of shock front distorsions, including a quantitative analysis of their departures from the 
nominal shock \tbn, have been recently reported for a perfectly perpendicular shock in fully three-dimensional geometry in \citet{Trotta2023c}.

\begin{figure}
\includegraphics[width=0.5\textwidth]{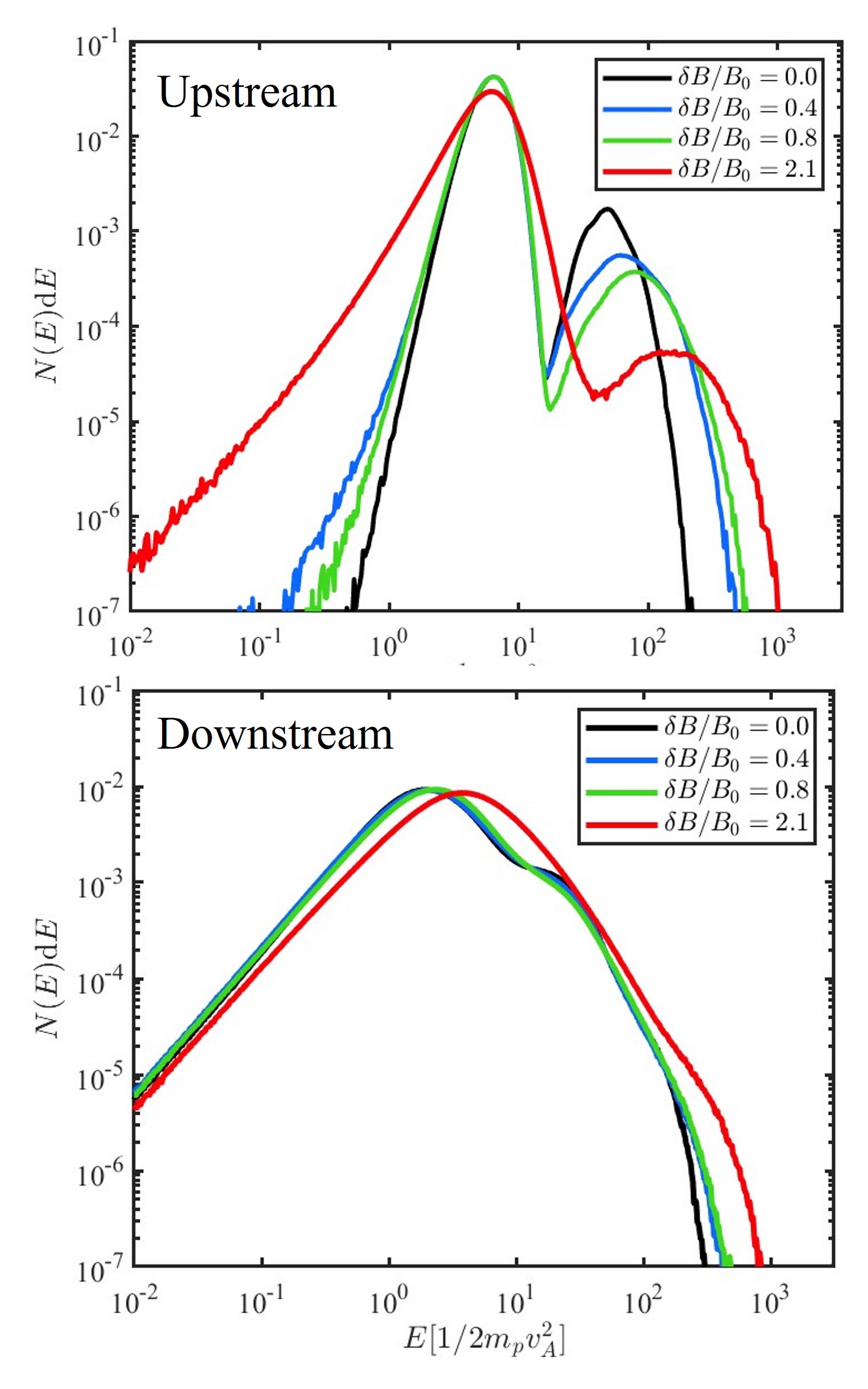}
\caption{\trotta{Upstream (top) and downstream (bottom) ion energy spectra for all the simulation cases, collected in 10 $\times$ 256 $d_i^2$ boxes upstream and downstream of each simulated shocks for the simulation times presented in Figure~\ref{fig:overview}. The colors represent the different runs with increasing turbulence levels}. \label{fig:spectra}}
\end{figure}

Another important feature resulting from Figure \ref{fig:overview} concerns the magnetic field patterns seen downstream. \trotta{Here, it is evident that when the shock propagates through a turbulent medium, there are strong downstream fluctuations not only in the vicinity of the shock transition but also further downstream, where many structures are observed due to the transmission of incoming turbulence, thus existing before the shock was formed and later processed by it. The study of such transmitted structures may have important implications as a promising pathway to further particle acceleration, as shown by recent theoretical~\citep{Zank2015}, numerical~\citep{Nakanotani2022} and observational works~\citep{Kilpua2023,Trotta2024a}.}.

Turbulent coherent structures are transmitted and distorted across the shock, and some other 
structures are created in the process (e.g., through magnetic reconnection between compressed magnetic islands,
see \citet{Gingell2023}). 
A complete analysis concerning the transmission of turbulent coherent structures and the properties of the perturbed shock 
downstream will be object of further investigations. Preliminary results, concerning exactly perpendicular shocks, have been 
recently reported in~\citet{Trotta2022a,Nakanotani2022}.

\trotta{The top panels of Figure \ref{fig:overview} show the one-dimensional magnetic field profiles along the $x$-axis for each simulation case. The $y$-averaged magnetic field profiles are shown (black lines) as well as two different one-dimensional horizontal cuts of the simulation domain taken at $y \, = \, 60 \, d_i$ and $210 \, d_i$ (red and blue lines, respectively). Two interesting considerations arise from analysing such one-dimensional profiles: first, in the unperturbed case, the one-dimensional magnetic field profiles are much closer to the $y$-averaged profile, indicating a higher level of coherence along the shock front. Conversely, when pre-existing turbulence is taken into account, strong departures from the averaged magnetic field profile can be observed. The second consideration is that, by comparing the $y$-averaged profiles for all the simulation cases, the shock transition appears less and less sharp. This is due to the stronger spatial distortions of the shock front for increasing levels of pre-existing turbulence. This analysis has important consequences for in-situ spacecraft investigations, crossing the shock in one point, underlining the need for multi-point observations at scales relevant to appreciate such shock front distorsions~\citep[see][]{Trotta2022b}.}

\trotta{The above analysis has been performed for fixed snapshots of the simulation domains, choosing times when the shocks were around the same position. However, time evolution plays an important role on shock front distortions and (irregular) production of energetic particles~\citep{Sundberg2016}. In Figure~\ref{fig:evolution}, we show different evolutionary stages of the shock front for three different simulation times, $\mathrm T \Omega_{ci} = 10,25,50$ (rows) and for all the simulation cases (columns with increasing pre-existing turbulence strength from left to right). In the unperturbed case, the shock transition starts completely laminar, and starts becoming distorted self-consistently due to the reflected particles creating unstable upstream distribution functions which, in turn, create fluctuations that are convected to the shock front. When pre-existing turbulence is present, the evolution of the system is completely different, with (expected) shock front distortions early on. Such disturbances, in turn, make it less favorable to have a homogeneous interaction of the upstream plasma with the shock front, potentially inhibiting the growth of self-consistently generated upstream fluctuations, with important implications on the observations of upstream fluctuations at both the Earth's bow shock~\citep{Eastwood2005} and at interplanetary shocks, where clear observations of foreshocks are often elusive and hard to disentangle from pre-existing turbulence~\citep[][]{Kajdic2012}. The interesting problem of studying the asymptotic behavior in time of shock-induced vs pre-existing fluctuations by employing longer simulation times and larger simulation domains will be object of further work, and it is out of this work's scope. Further, from Figure~\ref{fig:evolution}, it can be noted that for different times the shocks are found at different positions of the simulation domain, an effect due to the fact that with higher level of pre-existing turbulence we observe shocks with higher speeds, which are estimated to be of 1.4, 1.55, 1.7 and 2 $v_A$ in the negative $x$-direction in the downstream (simulation frame), respectively. The details about how these shock speeds have been computed can be found in Appendix~\ref{sec:appendix}.}

\trotta{Figure~\ref{fig:spectra} shows protons energy spectra, in the downstream rest frame,
for all simulation cases collected in 10 $\times$ 256 $d_i^2$ boxes upstream (top) and downstream (bottom) of the shock for the simulation times presented in Figure~\ref{fig:overview}.}. Both upstream and 
downstream, a population of accelerated particles is present together with the bulk population processed by the
shock. It is then possible to see 
that, when pre-existing turbulence is taken into account, several interesting effects take place. Higher maximum energies are observed with increasing levels of upstream fluctuations. Additionally, a low-energy tail of particles is particularly evident upstream, likely due to turbulent structures hindering particle motion. We note that the downstream spectra found in these simulations are compatible with the ones reported by \citet{Nakanotani2022} in a similar setup, employing a perpendicular shock 
propagating through turbulence. 

So far, we observed how the picture changes for a shock interacting with an increasingly strong pre-existing turbulence, namely showing increasingly distorted shock fronts and broader energy spectra. We now focus on why these changes are crucial from the particles' perspective, addressing their phase--space transport properties. We cast to the importance of understanding the possibility for shock-reflected particles to be scattered back to the shock front, where they may be further accelerated, fundamental ingredient of shock acceleration theories. To this end, we describe the details of a novel approach to study phase space diffusion, based on the coarse-graining of the Vlasov equation.

\section{The coarse-grained Vlasov Model}
\label{sec:particles_transport}
Typically, the investigation of particle transport properties requires a detailed analysis of energetic particles trajectories \citep[e.g.,][]{Caprioli2014c,Servidio2016,Comisso2019}. However, particle tracing is a technique that suffers several limitations, such as, for example, being inapplicable to direct observations \citep[e.g.,][]{Perri2015}. Furthermore, shock simulations are intrinsically non-periodic and strongly inhomogeneous, making it particularly difficult to follow large ensembles of particles for long enough times \citep[see][]{Trotta2020a}. 

Eulerian approaches to particles transport have a number of interesting advantages. An  example is the field-particle correlation technique \citet{Howes2017} analysing secular energy transfers between fields and particles, tested in numerical and observational setups \citep{Klein2017b,Chen2019, Klein2020}. 

We employ a novel Eulerian method to investigate phase space, which relies on the coarse-graining of the Vlasov equation. This method operates entirely within phase space, and is based on subsequent filtering techniques as described below.

The premise is that the shock-turbulence interaction is a multi-scale process characterized by a variety of ``lengths'', that in a collisionless (or weakly collisional) plasma involve both the physical and velocity sub-spaces.  In such multidimensional complexity, it is natural to employ coarse-graining and filtering techniques. Coarse-graining has been successfully applied in various areas of fluid dynamics and MHD, resulting in successful descriptions of a wide range of systems, from river flows to large-scale astrophysical systems~\citep[][]{Leonard1974,Alfonsi2019,Kaepylae2020,Vazza2014}. 
Recently, important progresses on coarse-graining and filtering approaches have been made also in the plasma kinetic framework~\citep[e.g.,][]{Eyink2018}, as well as in fluid plasma modeling \citep{Camporeale2018, Cerri2020}.

\begin{figure}
\includegraphics[width=0.48\textwidth]{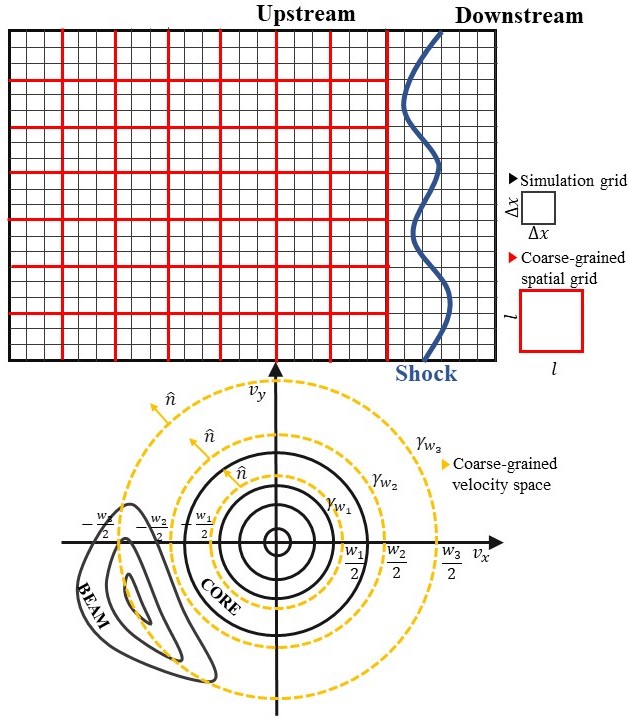}
\caption{Schematics of the coarse-graining approach. \emph{Top:} Sketch of the simulation domain, where the simulation 
grid and the coarse-grained grid of spatial extent $l$ are highlighted. \emph{Bottom:} Schematic representation of ion 
VDFs in an upstream coarse-grained cell (black). Three different choices of $w$ and subsequent integration paths 
$\gamma_w$ are shown (yellow).\label{fig:cg_scheme}}
\end{figure}

The starting point of the coarse grained approach for phase space transport is the Vlasov equation:
\begin{equation}
	\label{eq:Vlasov_first}
    \partial_t f +  \mathbf{v} \cdot \nabla  f + \left(\mathbf{E} + \mathbf{v} \times \mathbf{B} \right) \cdot \nabla_v f  = 0. 
\end{equation}
Here, $f = f(\mathbf{x}, \mathbf{v},t)$ is the protons' velocity distribution function (VDF), $\mathbf{E}$ and $\mathbf{B}$ are the electric and magnetic 
fields, respectively, and $\partial t$, $\nabla$ and $\nabla_v$ are the time derivative and the gradient operators in 
physical and velocity spaces, respectively.

Given the reduced dimensionality of our simulations and in order to simplify the model at a basic level, we reduce to 
a 2D-2V description. By integrating along $v_z$, we define $F(x,y,v_x,v_y,t) = \int_{-
\infty}^{+\infty}f(\mathbf{x},\mathbf{v},t)  \mathrm{d} v_z$. Through integration along $z$ and $v_z$, Equation 
\ref{eq:Vlasov_first} becomes: 
\begin{equation}
	\label{eq:Vlasov2}
    \partial_t F + \nabla \cdot \left( \mathbf{v} F \right) + \nabla_v \cdot \left( \mathbf{P} F \right) = 0, 
\end{equation}
where ${\bf E}$ and ${\bf B}$ have been incorporated in the term $\mathbf{P} = \mathbf{P}({\bf x}, {\bf v})={\bf E}+{\bf v}\times{\bf B}$, on reduced 4D phase space with coordinates ($x, y, v_x, v_y$).

Similarly to \citep{Eyink2018}, we obtain the coarse-grained Vlasov equation by defining a scale-dependent, filtered distribution
\begin{equation}
    \overline{F}_l(\mathbf{x}, \mathbf{v}, t) = \int_{-\infty}^{+\infty} f(\mathbf{x} + \mathbf{r}, \mathbf{v}, t) G_l(\mathbf{r}) d^2 r, 
\end{equation}
where $G_l(\mathbf{r})$ is a kernel that satisfies a series of properties, being non-negative, normalized and centered, and rapidly approaching to zero for $r\rightarrow \infty$. In our case we chose the simplest box-filter type, that in the reduced 2D Cartesian coordinate is given by $ G_l(\mathbf{r}) = 1/l^2$ for $|r_x|<l/2$ and  $|r_y|<l/2$, and equal to zero otherwise. 

The filter $G_l$ is then applied to Equation \ref{eq:Vlasov2}, yielding to:
\begin{equation}  
\label{eq:CGV1}
    \partial_t \overline{F}_l + \nabla_l \cdot \left[ \mathbf{v} \overline{F}_l \right] + \nabla_v \cdot \left[ \overline{\mathbf{P}}_l \overline{F}_l \right] = \nabla_v \cdot \overline{\mathbf{Q}}_l,
\end{equation}
where $\overline{\mathbf{P}F}_l = \overline{\mathbf{P}}_l \overline{F}_l - \overline{\mathbf{Q}}_l$. The latter decomposition, typical of Reynolds-averaging techniques \citep{Germano1991,Usmanov2018}, introduces the ``closure problem'', related to the description of the subgrid modeling \citep{Yang2018, Servidio2022}. This is a turbulent diffusion due to small scale fluctuations. At this point, Equation \ref{eq:CGV1} provides a kinetic description of the system in a coarse-grained physical space, where there is a residual flux related to the small scale correlation between fields and particle VDFs, and $l$ is coarse graining scale.

As discussed above, it is now necessary to deal with velocity space. In analogy with the Parker equation for energetic particles transport, we choose a typical speed $w$ (and therefore a typical energy) at which we integrate Equation~\ref{eq:CGV1} \citep{Jokipii1987}. The choice of $w$ is arbitrary and it depends from which portion of velocity space we desire to focus (sub-thermal, thermal or supra-thermal).

\begin{figure}
\includegraphics[width=0.44\textwidth]{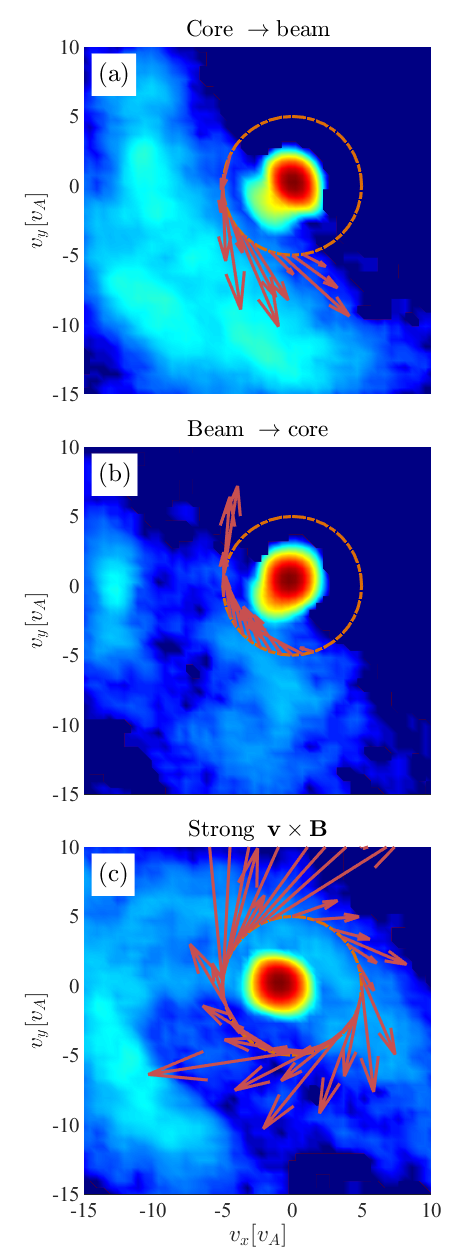}
\caption{Examples of coarse-grained VDFs and fluxes in phase space. The dashed line represents the velocity--space circle $\gamma_w$, where a radius $w= 5 v_A$ was chosen. Arrows represent $\bar{F_l}\bar{\bf{P}_l}$ along the circle. a) and b) show two cases with postive and negative flux, while case c) represents a case where the net flux is zero. \label{fig:vdf_flux}}
\end{figure}

At this point, for a chosen $w$, it is possible to define the reduced moments of the coarse-grained Vlasov equation in a similar fashion to how the Vlasov equation moments are defined \citep[e.g.][]{Krall1973}.
For instance, the zeroth-reduced moment \citep{Parker1965,Jokipii2010}, namely the reduced density, becomes
\begin{equation}
\overline{N}_{l, w}(x,y,t) = \int d^2 v \overline{F}_l(x, y, {\bf v}, t) G_w({\bf v}),
\end{equation}
where now $G_w({\bf v})$ is a unitary kernel, different from zero only for $|{\bf v}|<w$ and $\overline{N}_{l, w}(x,y,t)$ is the thermal population, coarse-grained in space at scale $l$. 

After multiplying by $G_w({\bf v})$ and by integrating Eq.~\ref{eq:CGV1}, one gets
\begin{equation}
	\label{eq:CGV2}
    \partial_t \overline{N}_{l,w} + \nabla \cdot\left[ \overline{N}_{l,w} \overline{\mathbf{V}}_{l,w}  \right] 
    + \int_{\gamma_w}\!\!\!\!\overline{F}_l \overline{\mathbf{P}}_l \cdot \hat{\mathbf{n}}\,d\gamma 
    = \int_{\gamma_w}\!\!\!\!\overline{\mathbf{Q}}_l \cdot \hat{\mathbf{n}}\,d\gamma.
\end{equation}

The first term represents the time variation of the reduced density, the second term is responsible for spatial transport of particles over the coarse-grained space, while the third is a surface integral around the circle $\gamma_w$ of radius $w$. The right-hand side is the residual contribution from the subgrid scales.

\begin{figure}
	\centering
	\includegraphics[width=0.48\textwidth]{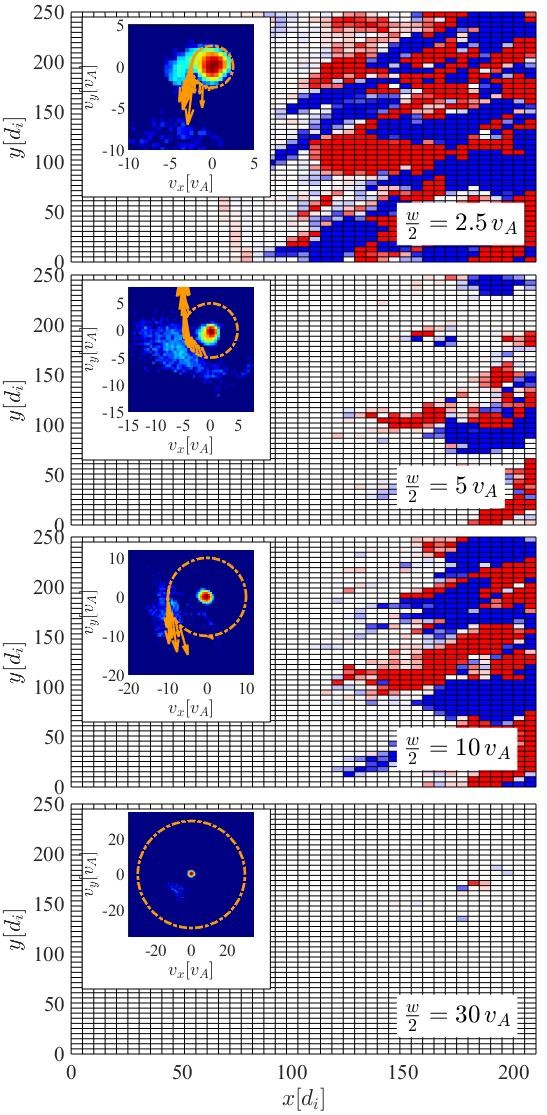}
	\caption{Coarse grained velocity space transport term $\int_{\gamma_w}\!\!\!\!\overline{F}_l \overline{\mathbf{P}}_l \cdot \hat{\mathbf{n}}\,d\gamma$ performed choosing $l = 5 \, d_i$, with increasing $w$ (top to bottom), for the moderately perturbed case $\delta B/B_0 = 0.8$. The colormap is chosen to be red for positive flux, blue for negative flux, and white for zero net flux. The insets show examples proton VDFs collected in the coarse grained cell, together with the integration circles $\gamma_w$ and the fluxes across them. In the insets, the velocity space transport term is positive, negative, positive and null from top to bottom.  
	\label{fig:w_new}}
\end{figure}

Interestingly, the velocity-space propagation term of Equation \ref{eq:CGV2} is a flux across a surface identified by $w$ and is essentially due to the normal component of the electric field, while the magnetic part of $\overline{\mathbf{P}}_l$ is tangent to the surface, acting as a pitch-angle spreader \citep[e.g.][]{Lyons1974,Shalchi2020}. Equation \ref{eq:CGV2} results from the 2D divergence theorem in velocity space for plasmas, for spatial integration at a length-scale $l$ and at a velocity cutoff $w$. Figure~\ref{fig:cg_scheme} represent a schematic of the approach, referenced though the work.

\begin{figure*}
	\centering
	\includegraphics[width=0.85\textwidth]{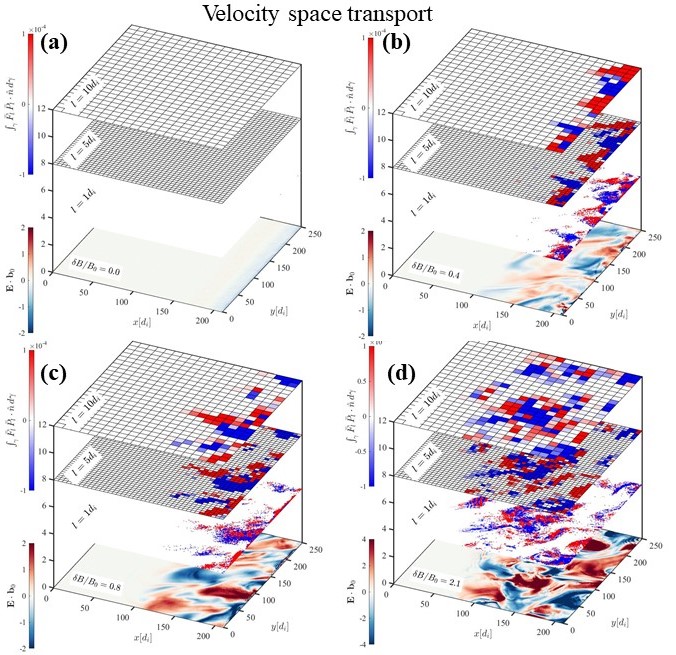}
	\caption{Vertical stack of plots showing upstream velocity space transport for all four cases $\delta B/B_0 \sim 0, 0.4, 0.8, 2.1$ (a-d, respectively). In each stack, the bottom panel shows a colormap of upstream parallel electric field $\bf{E} \cdot \hat{\bf{b}}_0$, while the velocity space transport term $\int_{\gamma_w}\!\!\!\!\overline{F}_l \overline{\mathbf{P}}_l \cdot \hat{\mathbf{n}}\,d\gamma$ is shown in the top three plots of the stack for different choices of $l$ ($l = 1,5,10 d_i$, respectively). As in Figure~\ref{fig:w_new}, the colormap is chosen to be red for positive flux, blue for negative flux, and white for zero net flux. In all cases, we chose $w = 10 \, v_A$, corresponding to a $\gamma_w$ circle passing between the inflow and beam population.\label{fig:velocity_pillars}}
\end{figure*}

\begin{figure*}
\centering
\includegraphics[width=0.85\textwidth]{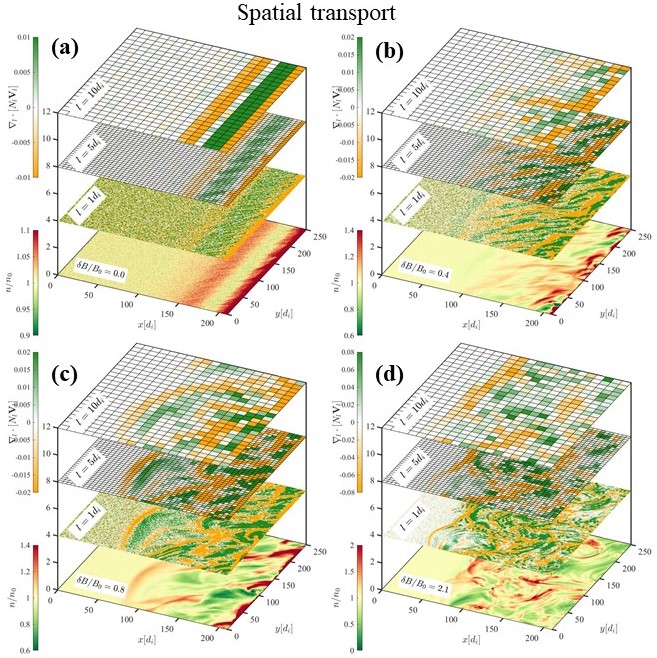}
\caption{Vertical stack of plots showing upstream velocity space transport for all four cases $\delta B/B_0 \sim 0, 0.4, 0.8, 2.1$ (a-d, respectively). In each stack, the bottom panel shows a colormap of upstream proton density, while the spatial transport term $ \nabla_l \cdot\left[ \overline{N}_{l,w} \overline{\mathbf{V}}_{l,w}  \right]$ is shown in the top three plots of the stack for different choices of $l$ ($l = 1,5,10 d_i$, respectively). In all cases, we chose $w = 10 \, v_A$, corresponding to a $\gamma_w$ circle passing between the inflow and beam population. \label{fig:spatial_pillars}}
\end{figure*}

\section{Ion phase--space transport in the transition from laminar to turbulent shock}
\label{subsec:CG_applied}

We now apply the coarse--graining technique to ion diffusion upstream of the simulated shock 
with and without pre-existing turbulence. The model's advantage lies in its ability to describe large-scale patterns of acceleration and diffusion processes. A simple examination of flux signs provides unique insights into spatial diffusion/clustering and energization/deceleration, as we will elaborate on later. For the case of oblique shocks propagating thorugh turbulence, a natural choice for the coarse-graining parameters $l$ and $w$ would be to select $l$ within the inertial range of upstream pre-existing turbulence and $w$ in such a manner that it delineates the inflow and FAB beam population (refer to $\gamma_{w2}$ in Figure \ref{fig:cg_scheme}). 

An example of the technique is reported in Figure \ref{fig:vdf_flux}, where three different cases for velocity space 
transport are shown through the study of $\overline{F}_l \overline{\mathbf{P}}_l$ for the $\delta B/B_0 \sim 0.8$ simulation case, where $l$ and $w/2$ were
fixed at 5 $d_i$ and $5 v_A$ (between the core and the FAB population in the unperturbed case), respectively. The 
study of the $\overline{F}_l \overline{\mathbf{P}}_l$ flux identifies three zones where transport in velocity space is
active in three different realisations, namely a region corresponding to particle acceleration from the core to the beam,
a region of deceleration (core to beam) and finally a region where transport is active with a strong $\mathbf{v} \times \mathbf{B}$ but zero net flux, thus scattering particles with no changes to their energies (Figure~\ref{fig:vdf_flux}a,b,c, respectively). It is important to note that all three realisation coexist in the upstream of the perturbed shock.

We now study phase--space fluxes in the perturbed shock upstream systematically. A crucial feature of collisionless plasmas is their ability to generate distributions that are far from equilibrium in velocity space. As demonstrated above, by applying coarse-graining to the Vlasov equation, a typical speed (energy) $w$, at which the velocity space integration is performed, can be selected. For the case of the oblique shock upstream, by choosing different values of $w$, it is possible to investigate various portions of velocity space, corresponding to distinct features of proton VDFs, such as inflow and beam components.

In Figure \ref{fig:w_new}, we show the coarse grained velocity space transport term 
$\int_{\gamma_w}\!\!\!\!\overline{F}_l \overline{\mathbf{P}}_l \cdot \hat{\mathbf{n}}\,d\gamma$ for a fixed $l$ (5 
$d_i$), and for four cases with increasing values of $w$. This analysis is performed on the moderately perturbed shock simulation, with $\delta B/B_0 = 0.8$. The values of $w$ have been chosen to be such that $\frac{w}{2}$ is 2.5, 5, 10 and 30 
$v_A$, corresponding to weakly suprathermal, inflow - beam gap, beam and high energy portions of the velocity space. This highlights how the methodology allows us to 
address different regions of velocity space where mechanisms of acceleration/deceleration 
are operating.

The spatial coarse-graining at scale $l$ yields this ``mosaic'' representation of the shock upstream where regions of positive, negative, and zero net flux are readily identified (red, blue and white in Figure~\ref{fig:w_new}).

When $w/2=2.5 v_A$, velocity space transport is highly active due to the effects of turbulence on particle acceleration and cooling. As we shall see later (Figure~\ref{fig:velocity_pillars}), this region is activated only when pre-existing turbulence is present, thus establishing a ``bridge'' between the inflow and the FAB populations.

With $w/2=10 v_A$, the integration path is designed to pass through the FAB component of the typical upstream VDFs, which represents the most populated suprathermal area of the velocity space (see orange dashed lines in the figure insets). Here, we find that velocity space transport is very strong, in particular close to the shock transition, 
due to particle scattering that is internal to the beam and enhanced by turbulent fluctuations. Finally, when a large 
$w$ parameter is chosen (Figure \ref{fig:w_new}, bottom plot, $w/2 = 30 \, v_A$), the velocity space transport 
term is switched off, due to absence of such high-energy particles.

The insets of Figure \ref{fig:w_new} show examples of fluxes in velocity space in the coarse grained cells, carried out in a similar fashion as Figure~\ref{fig:vdf_flux}, where the 
vector $\overline{F}_l \overline{\mathbf{P}}_l $ is shown on the integration path (arrows). From top to bottom, insets show positive (acceleration), negative (deceleration), positive, and zero net flux (no active transport) through $\gamma_w$ for the corresponding choices of $w$.

We now restrict the analysis of velocity space transport to $w=10 v_A$, highlighting the portion of velocity space between the inflow and FAB population for all simulation cases. Generally speaking, this is a unique identification method for places where injection of accelerated particles occurs.
Figure~\ref{fig:velocity_pillars} shows the outcome of this analysis, where we have chosen 3 different values of spatial coarse-graining length scale $l = 1,5,10 d_i$, respectively (see stacked mosaics in the Figure, with the same colorbar as in Figure~\ref{fig:w_new}).

Figure~\ref{fig:velocity_pillars} shows both how turbulence enhances phase space transport and why it is important to address it with our eulerian approach. In the unperturbed case $\delta B/B_0 = 0.0$, we find the velocity space transport term to be negligible at all scales 
$l$. This depends on the fact that, in the unperturbed case, the inflow and beam population are well-separated, and  the choice of the integration circle $\gamma_w$ with $w = 10 \, v_A$ results in a negligible flux. Thus, from the point of view of phase space transport, inflow and beam population may be considered as ``non-interacting''.

It can be seen from Figure~\ref{fig:velocity_pillars} that upstream effective phase space transport gets activated in the transition from laminar to turbulent shock. The heightened velocity fluxes clearly result from the intricate interaction between particles and turbulent fields. Here, the ${\bf v}\times{\bf B}$ force serves as a pitch angle spreader across the surface, while the turbulent electric field locally amplifies momentum diffusion. This amplification may happen through various local processes, including wave-particle interactions, as well as linear and non-linear Landau damping \citep{Valentini2005, Valentini2005b, Howes2017, Chen2019}, stochastic ion heating \citep{Chandran2013} and 
possible interaction with reconnetion processes in the upstream turbulent layer \citep{Zank2015,Servidio2009,Franci2017, Howes2017}.

We find a good correlation between the coarse-grained velocity space transport
quantities and the upstream parallel (turbulent) electric field $\bf{E} \cdot \hat{\bf{b}}_0$, as shown in the bottom 
plots of Figure \ref{fig:velocity_pillars} and originally reported in \citet{Trotta2021}. Large fluxes and parallel electric field are, as expected, anti-
correlated, suggesting that any positive parallel electric field energize particles. On the other hand, the term 
$\bf{v} \times \bf{B}$ in the acceleration term $\overline{\bf{P}}_l$ of Equation \ref{eq:CGV2}, is tangent to the 
integration circle in velocity space $\gamma_w$, and acts as a pitch angle spreader, with no net contribution of the 
flux across the circle \citep{Lyons1974}.

Crucially, when examining the coarse-grained representation of the spatial transport term with varying $l$ parameters, it becomes evident that a crucial characteristic emerges, i. e. the presence of a self-similar behaviour. This property is very important since is suggests that there is some cross-scale similarity for the transport properties that makes our analysis valid in all the inertial range scales of turbulence. Further, the self-similarity recovered for different spatial coarse-graining scales is an important indicator of the robustness of our diagnostic. 

For completeness, we discuss results addressing the spatial transport term for the shock upstream in all simulation cases. The spatial transport term provides us with information about regions of spatial dilatation or compression  for the reduced density. 

Figure \ref{fig:spatial_pillars} shows results for the spatial transport 
term  $ \nabla_l \cdot\left[ \overline{N}_{l,w} \overline{\mathbf{V}}_{l,w}  \right]$, 
using different values of $l$ ($l = 1,5,10 d_i$) in the coarse graining, while keeping $w$ fixed at 10 $v_A$, as done for the analysis of velocity space transport in Figure~\ref{fig:velocity_pillars}, i.e., at the interface between inflow and FAB populations. Due to this choice of $w$, the reduced proton density $\overline{N}_{l,w}$ 
can be seen as the thermal particles' density (see Figure \ref{fig:cg_scheme}, $\gamma_{w2}$ for reference). 
Thus,  when the space-transport is positive, thermal particle are escaping, in the sense that their density is decreasing, and we are in presence of a sink of thermal particles. On the other hand, when the spatial transport term is negative, the thermal density is increasing, and plasma condensates. 

Figure~\ref{fig:spatial_pillars} yields important information about spatial transport in the shock upstream in its transition from laminar to turbulent. First of all, with 
increasing upstream turbulence strength, the spatial transport term  turns out to be enhanced. Transitioning from the unperturbed (Figure \ref{fig:spatial_pillars}a) to the perturbed (b-d) cases, we observe a loss of ``organisation'' in the spatial transport patterns due to turbulence.

The bottom plots of Figure \ref{fig:spatial_pillars} display upstream proton density, showing that there is a similar modulation with the spatial transport term. This is consistent with the fact that the reduced density is mostly made of thermal particles. This is evident in Figure \ref{fig:spatial_pillars}, the spatial counterpart of that can also be observed in velocity space in Figure \ref{fig:velocity_pillars}.
 
 \trotta{The self-similarity of coarse-grained transport quantities holds significant importance, as the contributions from spatial and velocity space (left hand contributions in Equation~\ref{eq:CGV2}) are the predominant factors. The residual component, represented by  $\int_{\gamma_w}\!\!\!\!\overline{\mathbf{Q}}_l \cdot \hat{\mathbf{n}}\,d\gamma$  in the right-hand side of Equation~\ref{eq:CGV2}, consists of small-scale contributions that are markedly intermittent and comparatively negligible with respect the other terms (not shown here). Furthermore, the small scale residual is more difficult to be estimated, because of both resolution limitations and particle noise of the PIC model.}
Finally, we underline that the $l$ chosen in our analysis span the inertial range of upstream turbulence, and the self-similar behaviour is expected on arguments inspired by the turbulent nature of 
upstream fluctuations. We remark that this self-similar behaviour is also typical of inertial range coarse-graining in fluids \citep[e.g.,][]{Frisch1995,Lesieur1996}.

\section{Conclusions}
\label{sec:conclusions}

In this work, we focused on the transition from a laminar to a turbulent oblique shock by performing hybrid kinetic simulations of the shock-turbulence interaction. The simulations, firstly introduced in \citet{Trotta2021}, represent a step forward in describing shock propagation in realistic turbulence down to ion kinetic scales. \trotta{This study has two key results: first, it shows  the shock behaviour changes when pre-existing turbulent is taken into account, then, it elucidates how shock-reflected particles diffuse in the shock upstream with and without pre-existing turbulence, thereby showing how phase-space diffusion is enhanced for higher levels of upstream turbulence.} Our approach, useful to study the early-time evolution of the system, is complementary to other approaches recently adopted to study different aspects 
of the shock turbulence interaction~\citep[e.g.][]{Nakanotani2022,Behar2022}.

We first demonstrated how various shock properties change with different levels of upstream turbulence ($\delta B/B_0 \sim 0, 0.4, 0.8, 2.1$), consistent with previous studies conducted in both two-dimensional and fully three-dimensional geometries~\citep{Trotta2022a,Nakanotani2022,Trotta2023c}. In 
particular, upstream and downstream proton energy spectra have been found to be affected by pre-existing turbulence, with higher energies achieved for higher levels of turbulent fluctuations. Such results can be put in the context of observations of shock-accelerated particles at interplanetary shocks, with different levels of upstream fluctuations~\citep[see, for example][]{Lario2022,Trotta2023b}, as well as in the context of shock 
reflected particles upstream of Earth's bow shock~\citep{Kucharek2004}. 

We then showed how phase space transport is activated when upstream turbulence is present. Using a novel Eulerian approach to address diffusion, which relies on the coarse-graining of the Vlasov equation, we could address a critical portion of velocity space between inflow and shock reflected particles. The coarse-graining approach allowed up to highlight how turbulence forms a ``bridge'' between these two population, that interact and mix upstream in all perturbed cases. \trotta{As expected, these two different particle populations, namely the upstream inflow and reflected particles, have been found to mix more effectively for stronger levels of pre-existing turbulence, here shown by means of novel, self-consistent simulations.}

The novel coarse-graining methodology has been thoroughly discussed. Here, we combine spatial filtering, typical of hydrodynamics, with Parker-type transport equations, typical of cosmic ray 
physics \citep[e.g.,][]{Amato2017}. By averaging over inertial range scales and applying the divergence theorem in velocity space, the turbulent upstream can be described as a ``mosaic''. Each piece of this mosaic is characterized by strong spatial dilation and condensation, driven by Alfvénic turbulent modulations. More interestingly, the velocity-space filtered flux is noticeably anti-correlated with the parallel electric field, suggesting the possibility \trotta{of various field-particle interactions, for which a complete characterisation is out of scope for this paper and will be part of future investigations}. The results on the phase space transport yield self-similar results for spatial scales in the turbulence intertial range, a crucial indication of robustness of the diagnostic that has implications on its applicability to spacecraft measurements.

\trotta{Our coarse grained analysis is limited to the spatial filtering, while in the velocity scale we proceeded with a cutoff. This was necessary in order to understand the injection problem and the “transfer” between the thermal parte and the suprathermal field-aligned beam. A coarse graining in the velocity scale would be inherently related to the phase space cascade of entropy, from large to finer phase space scales, invoked in \citet{Eyink2018, Servidio2017, Schekochihin2016}. Although this direction is also interesting (and will be explored in future investigations), they point toward another meaning, different from the present investigation. In our case the procedure is similar to the Parker-transport equation from the velocity side, and to the Germano techniques regarding the spatial analyses \citep{Yang2016}.}

These results might have important consequences on the understanding of transport and heating processes in a variety of space and astrophysical systems. However, this work has important limitations, such as the reduced dimensionality employed, as well as the fact that electron kinetics has been neglected (for computational reasons). 

In future works, we will extend the analysis to in-situ measurements, where the possibility of choosing a coarse-grained spatial scale could relax the time-resolution constraints of spacecraft measurements. From this point of view, this work can be put in the context of application to direct observations of other important diagnostics, such as, for example the Hermite decomposition of plasma distribution functions \citep{Servidio2017, Pezzi2018}.

\trotta{Another important avenue for further investigations will be to include variable upstream conditions, for example forcing the pre-existing turbulence amplitude to grow/decay in time, or to study the convection of solar wind structures such as flux ropes embedded in turbulent media~\citep{Pezzi2024} across the shock and addressing their role in particle acceleration.}

\section*{Acknowledgements}

This work has received funding from the European Unions Horizon 2020 research and innovation programme under grant agreement No. 101004159 (SERPENTINE, www.serpentine-h2020.eu) and from the European Union’s Horizon Europe research and innovation programme under grant agreement No. 101082633 (ASAP). Views and opinions expressed are however those of the authors only and do not necessarily reflect those of the European Union or the European Commission. Neither the European Union nor the European Commission can be held responsible for them.
FV acknowledges the support of the PRIN 2022 project “2022KL38BK - The ULtimate fate of TuRbulence from space to laboratory plAsmas (ULTRA)” (Master CUP B53D23004850006; CUP H53D23000930006) by the Italian Ministry of University and Research, ‘Finanziato dall’Unione europea – Next Generation EU’ PIANO NAZIONALEDI RIPRESA E RESILIENZA (PNRR) Missione 4 “Istruzione e Ricerca” - Componente C2 Investimento 1.1, ‘Fondo per il Programma Nazionale di Ricerca e Progetti di Rilevante Interesse Nazionale (PRIN)’ Settore PE9.
SS acknowledges the support from ICSC-Centro Nazionale di Ricerca in High Performance Computing, Big Data and Quantum Computing-and host-ing entity, funded by European Union-NextGenerationEU.
DB acknowledges support from UK Science and Technology Facilities Council (STFC) grant ST/X000974/1.

\section*{Data Availability}
The simulation datasets used for the analyses in this work can be found and freely downloaded here: 10.5281/zenodo.13730180. The authors will share further datasets from the simulations upon request.



\bibliographystyle{mnras}




\appendix

\section{Shock speed estimation}
\label{sec:appendix}
\begin{figure*}
\centering
\includegraphics[width=0.99\textwidth]{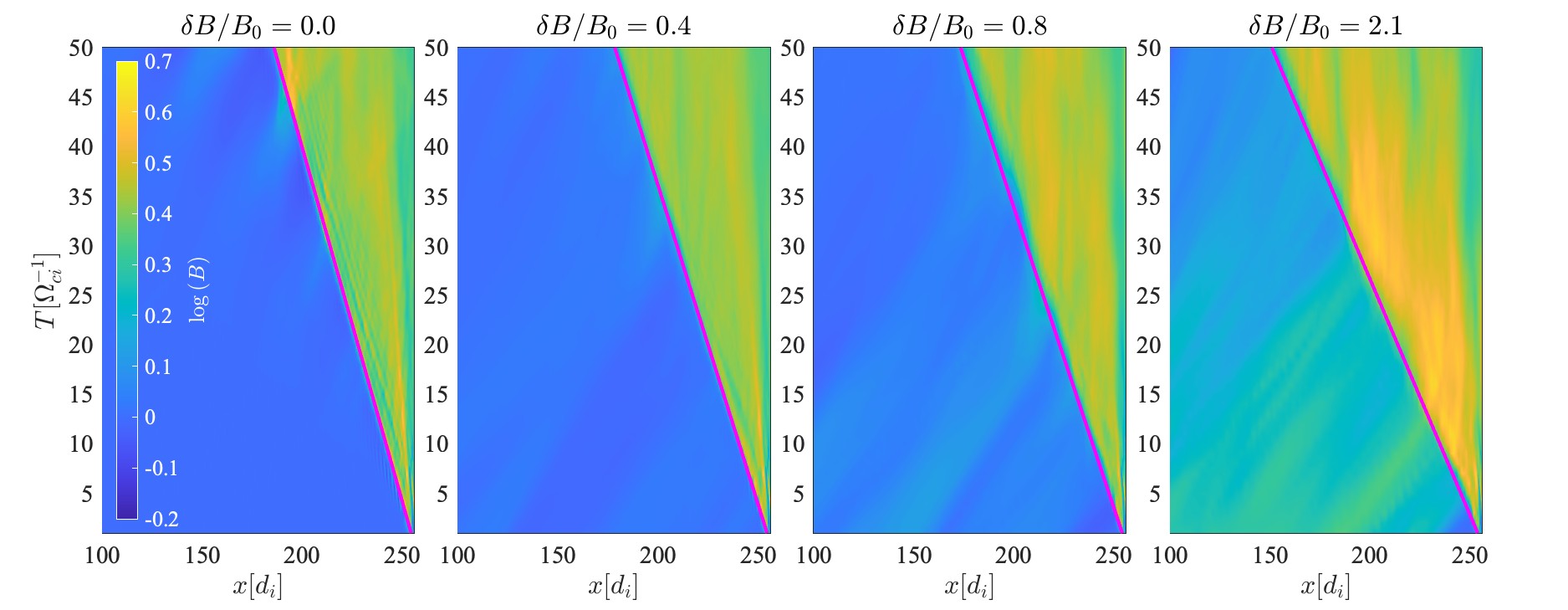}
\caption{\trotta{Colormaps showing simulation time vs $y$-averaged magnetic field along the $x$-direction for all the simulation cases (left to right). The magenta lines represents the nominal shock position in time. } \label{fig:speeds}}
\end{figure*}

\trotta{In this Appendix, we provide details about how the shock speed is computed in the simulations. We remind the reader that the simulate shocks are generated using the injection method~\citep{Quest1985}, and thus the simulations are performed in the downstream rest frame. The inflow speed $v_{in}$ is a free parameter used for initialise a supersonic (superalfvenic) plasma flow, and the shock speed is therefore not known \textit{a priori}. The first step of the procedure is to identify, for each simulation time, a (nominal) shock position. To this end, we study the $y$ averaged magnetic field profile in the $x$-direction of the simulation and locate the magnetic field jump identifying the shock transition for each time. In this way, it is possible to generate a two-dimensional map of magnetic field intensity as a function of space and time. Such maps are shown for all the simulation cases in Figure~\ref{fig:speeds}, with increasing turbulence levels from left to right.
At this point, the nominal shock position $x_{\mathrm{sh}}$ in time can be identified (magenta lines in Figure~\ref{fig:speeds}), and thus a shock speed can be simply derived as $v_{\mathrm{sh}} = \frac{d x_{\rm{sh}}}{d t}$. The results, for each simulation case, are shown in Table~\ref{tab:speeds}, where is can be seen that larger pre-existing turbulence levels are associated with higher shock speeds. However, we note that shock speed can exhibit strong departures from its mean estimation both in time and across different shock locations in space.}

\begin{table}
	\centering
	\caption{Shock speed estimations for each simulation case.}
	\label{tab:speeds}
	\begin{tabular}{lcccr} 
		\hline
		$\delta B/B_0$ & 0.0 & 0.4  & 0.8 & 2.1\\
		\hline
		$v_{\mathrm{sh}}$ & 1.4 & 1.55  & 1.7  & 2\\
		\hline
	\end{tabular}
\end{table}



\bsp	
\label{lastpage}
\end{document}